\documentclass[10pt,preprint]{aastex}

\slugcomment{To appear: ApJS, v161 (Nov), 2005}
\lefthead{Hanson et al.}
\righthead{NIR Spectral Atlas}

\begin{document}  

\title{A Medium Resolution Near-Infrared Spectral Atlas of O and Early B Stars}
\author{M.M. Hanson$^{1,5}$, R.-P. Kudritzki$^2$, M.A. Kenworthy$^{1,3}$}
\author{J. Puls$^4$, A.T. Tokunaga$^2$}

\altaffiltext{1}{Department of Physics, The University of Cincinnati, Cincinnati, 
OH 45221-0011}
\altaffiltext{2}{Institute for Astronomy, University of Hawaii, 2680 Woodlawn Drive, Honolulu, HI 96822}
\altaffiltext{3}{Present Address: Steward Observatory, University of Arizona, 933 Cherry Avenue, Tucson AZ 85721-0065} 
\altaffiltext{4}{Universit\"ats-Sternwarte M\"unchen, Scheinerstr. 1, D-81679 M\"unchen, Germany}
\altaffiltext{5}{Visiting astronomer, The Subaru Observatory and The Very Large Telescope}

\begin{abstract}

We present intermediate resolution (R $\sim$ 8,000 - 12,000) high signal-to-noise $H-$ and $K-$band spectroscopy
of a sample of 37 optically visible stars, ranging in spectral type from O3 to B3 and representing most
luminosity classes.   Spectra of this quality can be used to constrain
the temperature, luminosity and general wind properties of OB stars, when used in conjunction with 
sophisticated atmospheric model codes.  Most important is the need for
moderately high resolutions (R $\ge 5000$) and very high signal-to-noise (S/N $\ge$ 150) spectra 
for a meaningful profile analysis. 
When using near-infrared spectra for a classification system, moderately high signal-to-noise 
(S/N $\sim 100$) is still required, though the resolution can be relaxed to just a thousand or two.
  In the appendix we provide a set of very 
high quality near-infrared spectra of Brackett lines in six early-A dwarfs.  These can be used 
to aid in the modeling and removal of such lines when early-A dwarfs are used for 
telluric spectroscopic standards.

\end{abstract} 


\section{INTRODUCTION}  

  Kleinmann \& Hall (1986) were the first 
to present reasonably-high resolution, high signal-to-noise (S/N) near-infrared 
(NIR) spectra for cool stars.  The first NIR spectral atlas of hot stars
was given by Lan\c{c}on \& Rocca-Volmerage (1992). Designed for use in stellar
population synthesis models, the Lan\c{c}on \& Rocca-Volmerage atlas 
lacked adequate resolution for applications in many stellar and 
galactic programs.  A few years later, Dallier et al.\ (1996) and
Hanson et al.\ (1996) presented $H$-band and $K$-band spectral atlases, respectively,
which included OB stars with significantly higher resolution and S/N. 
Numerous NIR atlases of OB stars have been published since that time 
(Wallace \& Hinkle 1997, Lenorzer et al.\ 2002, for a recent review of all NIR spectral 
atlases, see Ivanov et al.\ 2004). The utility of a NIR spectral classification scheme, 
for hot stars in particular, has proved exceedingly useful for a variety of applications, 
including studies of very young star forming regions (Bik et al.\ 2003) and
the galactic center region (Najarro et al. 1997, Ghez et al. 2003; Najarro et al.\ 2004) 
as well as distant massive clusters 
through out the Galaxy (Hanson, Conti \& Howarth 1997, Blum, Damineli \& Conti 1999;
Figer et al.\ 2005). 
Furthermore, researchers studying heavily reddened high-mass X-ray binary systems 
(Clark et al.\ 2003; Morel \& Grosdidier 2005) and microquasars (Mirabel et al.\ 1997, Mart\`{i} et al. 2000) have found 
NIR spectral classification to be uniquely valuable.

In light of these successes, our group wishes to push NIR spectral studies
of OB stars to a new, more sophisticated level.   Our goal is to obtain new, 
higher resolution and S/N
NIR OB spectra to test and guide existing quantitative atmospheric 
models for OB stars in the NIR regime (Najarro et al.\ 1998; Kudritzki \& 
Puls 2000 and references therein).  In turn, once the atmospheric models have
been calibrated to properly predict stellar characteristics based on the 
NIR spectra of known, UV- and optically-studied stars, it is our hope 
that they may be used to provide accurate constraints to the characteristics
of stars observable only in the NIR.  This NIR atlas of well 
known, optically visible OB stars makes up the sample of high-quality spectra 
which are being used by our group for a successful NIR qualitative analysis 
(Repolust et al.\ submitted).

\section{OBSERVATIONS}

A list of the stars used for the survey, their position, and salient details of the observations
is given in Table 1.  When our observations were carried out, reasonably-high-resolution 
($R > 8000$) spectrometers which allowed for sufficient spectral coverage were only available 
on 8- and 10-meter class telescopes.  Our targets are exceedingly bright for such a large 
aperture system.  However, these observations are
absolutely necessary for the sake of developing and testing quantitative model atmospheres.
Furthermore, we needed to start with optically visible, well known O and early-B stars.  The OB sample was
selected to give reasonable coverage of the temperature and luminosity range of O and early-B
stars.  The temperature and luminosity range sampled is illustrated in the spectral type
versus luminosity class presentation given in Table 2.

There presently are no classification standards in the NIR.  Until sufficient numbers of OB 
stars have been observed in the NIR, it will be too soon to claim any star 
as a classification standard.  The stars selected in this survey are {\bf not} being 
promoted as standard stars for spectral classification. The targets for this program 
were selected based on entirely different criteria.  The OB supergiant stars observed with the 
VLT were obtained as part of a study of OB supergiants near the galactic center (Fickenscher, 
Hanson \& Puls, 2004). For ease of the observing run, this required them to be in the vicinity 
of the galactic center fields being observed. Most of the stars observed with Subaru 
were hand selected by one of us (J.P.). The Subaru sample was selected to cover a 
reasonable sampling of effective temperature and gravity, and most importantly, with the 
expressed desire to model their spectral profiles using modifications to the atmospheric code 
FASTWIND (Puls et al.\ 2005).  Most of the stars selected have already undergone significant 
previous spectroscopic modeling in the optical and UV, and did not show any 
serious irregularities in those analyzes.  The very high resolution of these observations 
are well beyond those typically used for classification in
the optical (R $\sim$ few thousand).  Spectra for the purpose of
classification, no matter the wavelength, are best obtained at more moderate resolutions
(see further discussion of this point in \S 5.2).

\subsection{The VLT-ISAAC Spectra}

The first of our data came from the Infrared Spectrometer and Array Camera (ISAAC).
The instrument is mounted at Nasmyth focus on 
the 8.2m Unit 1 telescope of the European Southern Observatory's (ESO) Very Large Telescope (VLT), 
located on Cerro Paranal in Atacama, Chile (Moorwood 1997).  ISAAC employs a Rockwell 
Hawaii 1024$^2$ array and a single grating. The resolution is set by the slit width.
In May and April of 2001, long slit (120'') $H-$ and $K-$band spectra were obtained for a number
of optically visible late-O and early-B supergiants.  All data for this program were obtained 
in the queue observing mode, spread out over about two months and using approximately 12 hours
of VLT queue time.  Typically 8 slit positions were obtained, with total 
on-source integration times of one to a few minutes.  A slit-width 
of 0.3'' was used, giving a spectral resolving power of $R \sim 10,000$ in the $H-$band and 
$R \sim 8,000$ in the $K-$band.  Three grating settings, 1.710, 2.085, and 2.166 $\mu$m,  
were used with ISAAC.

\subsection{The Subaru-IRCS Spectra}

Later that year, we obtained additional spectra at the 8.2-m Subaru Telescope,
operated by the National Astronomical Observatory of Japan (Tokunaga et al. 1998)
and located at the top of Mauna Kea in Hawaii.  We employed the Infrared 
Camera and Spectrograph (IRCS) installed at Cassegrain focus.  IRCS uses two 1024$^2$ ALADDIN arrays 
and offers a cross-dispersed echelle mode for high resolution work.  With the present 
arrays, the cross-dispersed mode does not allow for full spectral coverage, leaving unobserved 
spectral regions between the echelle orders.  For full coverage, two settings can be completed.
However, nearly all the important lines for our survey (except the $\lambda 2.0581 \mu$m HeI line)
were observed with a single grating setting.
 
To achieve the highest resolution, we used the most narrow slit setting (0.15'').  
This provided a resolution of about 0.5 \AA/pixel in $K$, and about 0.4 \AA/pixel in $H$.  
In practice, we achieved an approximate FWHM of 3.5 pixels, measured through the OH sky 
emission features, resulting in a final spectral resolution of approximately $R \sim 12,000$.
These spectra were obtained over two separate runs, first in November 2001, then later 
in July 2002, based on the targets' right ascension.  Two nights were granted
for each run.  The first night was used to obtain just the $K-$band spectra, 
the second night was reserved for the $H-$band spectra for the same stars.  
The weather in November 2001 was rather poor; clouds on the 
mountain had closed just about every other optical observatory.  Because our sources are
so bright (with $K$ magnitudes as large as 2), we managed to get sufficient counts
despite the weather. Telluric corrections did prove to be more problematic, 
however, because of the poor sky conditions.  Conditions in July 2002 were also less than
ideal, but again, given the brightness of our targets and the 8-m aperture we obtained
sufficient counts.  Intermittent heavy clouds did lead to some poor 
telluric corrections in the final spectra for some sources.  We failed to get a 
follow up $H-$ band spectrum of HD 15558 during the second night.  Also, the $H-$band
spectrum of $\tau$ Sco, HD 149757, was somehow corrupted. Regrettably, a trustworthy 
spectrum proved irretrievable from our raw data.

\section{REDUCTION OF THE SPECTRA}

All VLT-ISAAC spectra were reduced using {\tt IRAF}\footnote{IRAF is distributed by the 
National Optical Astronomy Observatories, which are operated by the Association of
Universities for Research in Astronomy, Inc., under cooperative agreement with the
National Science Foundation.} routines.  Subsequent analysis of the Subaru spectra 
was done using routines written in PerlDL\footnote{homepage at http://pdl.perl.org/}.

The OB stars and a few telluric standards (more on that presently) 
were observed with the traditional long-slit AB ``nodding'' pattern on both
telescopes (with a small, few arcsecond jitter to better
sample the detector).  This allows quick and effective
subtraction of background emission when the A slit position is subtracted from 
the B slit position several arcseconds away (and vice versa).  For all stars 
observed with the VLT, four AB pairs were 
obtained, giving us 8 individual spectra.  For the November 2001 Subaru run, 
three ABBA sets were used.  This was reduced to just 2 ABBA sets for the
July 2002 run.  Each two-dimensional spectral image was bias (and other instrumental
effects) subtracted before flat fielding.  For all spectra, regardless of 
telescope, dark frames and flat field frames were averaged together to form 
a master dark and flat frame. Then AB pairs are subtracted to remove
the last of the sky emission.  Finally, each individual spectrum from the set
were extracted, weighted and scaled before being averaged, using a 3-sigma rejection.  
Telluric OH emission lines (Rousselot et al.\ 2000), ubiquitous throughout 
the spectral range of our data, were used for wavelength calibrations.  

The Earth's atmosphere introduces a myriad of absorption lines into ground-based,
NIR spectra, and these need to be removed (Fig.\ 1).  Because the features are 
complex and continuously changing, {\it in situ} measures, via a telluric 
standard star, are the best means of constraining their character.  Owing to their
near featureless continuum (with the very big exception of the Brackett series
of Hydrogen), late-B and early-A dwarf stars are frequently used  in this 
regard.   For each target star observed with the VLT-ISAAC, a telluric 
standard star was observed either just before or after the target object.  
This telluric star was selected to exhibit the same airmass (to within
a few hundredths usually) as the target object, and to lie in the same
general sky direction of the target object.  AB nodding along the slit was used to 
observe the telluric stars. Identical procedures were used for the reduction 
of the telluric stars as was used for the OB stars as outlined above.  For the VLT run, we observed a 
late-B dwarf, matched to each OB target star and put into individual observing 
blocks (as required for VLT queue observing).   We also obtained spectra of a
few early-G dwarfs in some of the observing blocks.  These early-G dwarfs are used with
the NIR spectrum of the Sun to help constrain the Brackett features in
the B-dwarf star.  Once the Brackett features have been modeled in the late-B 
dwarf, they can be divided from the late-B dwarf spectrum to create the 
telluric spectrum.  The telluric spectrum is then finally divided from 
the raw OB target star spectrum (see the Appendix in Hanson et al.\ 1996).

For the Subaru run, observations were taken in standard visitor mode.
A number of early-A dwarfs were observed through out the night on the assumption
that they would be used in the removal of telluric features.  In the
end, however, they had very limited use.   For OB stellar lines
far from the influence of any Brackett series transitions (HeII, most of the
HeI lines, etc.) the A stars were useful to derive the telluric spectrum,
but the Subaru spectra had unprecedented S/N
and even higher resolution than the VLT-ISAAC spectra.  Consequently, any
estimated fit for the Hydrogen Brackett lines in the A dwarf stars could not
be adequately constrained to derive the OB star hydrogen lines.  Ironically, 
the most ideal telluric standards would be very hot stars, because they have 
the fewest, weakest lines of all normal stars.  In the end, we were faced with
using the OB stars themselves, taken throughout the night, as telluric 
standards for each other in deriving the Hydrogen Brackett lines. 
In the infrared, every spectrum is like an equation with two unknowns:
the telluric spectrum and the stellar spectrum.  Taking additional spectra
of other stars does not solve this, even if the telluric spectrum is the
same, since every new spectrum is a new equation, but adds an additional
unknown (the underlying spectrum of the new star).  This is the case ONLY
in the Brackett region because every star shows a unique  Brackett feature.  
At some point, one is forced to make an {\it assumption} about the 
Brackett line profile in one of the stars (it is
our experience that this is preferred over trying to make an assumption
about the strength and shape of the numerous telluric features that span 
the Brackett spectral range).  This reduces the number of unknowns to equal
the number of spectra (`equations') and allows for a solution.  Typically,
a Voigt profile was used as a first, reasonable guess for the most well
behaved stars observed during the night: late-O or early-B dwarfs.  This 
`solution' was then propagated
through all the stars observed that night, dividing one against another until
one finally resolves the underlying spectrum for all the stars from the
night.  If problems or inconsistencies are revealed, then a new assumption
can be made and propagated through the stars for that night.   

\subsection{The need for synthesized spectra}

Is there any way to know if our assumptions made to model our most 'normal'
stars are reasonable?  What is our likely error?  This would be very difficult
to determine if we had no idea of the expected profile shapes for our
OB stars.  Luckily, in conjunction (and as a parallel goal) with the development
of this high S/N atlas, our group is also expanding existing sophisticated
models of OB stellar atmospheres to predict and produce line profiles in
the NIR (Repolust et al.\ 2005).  Once a best guess was made and
the underlying spectrum of all the stars was completed, these were compared
with synthesized spectra our group created.  If our assumptions about the
Brackett profile are wrong, then very obvious patterns of inconsistencies
would appear in the comparison of data and models.  Indeed, it was found
that the Br10 line of Hydrogen appeared to be consistently too weak (as
compared to the model predictions) for all stars observed in November 2001.
We returned to the data, used a stronger line for the primary assumption,
and then this solution was propagated
through out the night.  We found that a more satisfactory fit was obtained for all
the star in this particular line after having done this.

But still, what is the expected error in our profiles?  Given the brightness 
of our target (and telluric) stars, flux is not so much a limit to our 
confidence (indeed we have very high S/N) it is the
accuracy of the telluric corrections and the assumptions made in fitting
the Brackett features.  For the November 2001 `inconsistency', the
strength of the Br10 line was increased by approximately 1.5\% compared
to the continuum. This made a significant difference in the Br10
line as seen in the O4 supergiant, HD 14947 (it was the weakness of the
Br10 line in HD 14947 which exposed the problem).  But such a 
small difference was essentially 
undetectable in the Br10 spectrum of the B2 dwarf, HD 36166.  Given the
goodness of the fits to the synthesized spectra and the reproducibility 
of the features using the telluric A-star standards, we expect the HeI and 
HeII lines to be good to at least 0.5\% of the continuum.  The Brackett lines 
pose special problems, as we've outlined.  We estimate those lines to be good 
to between 0.5\% and 1\% through out the wings, and perhaps only good to 
1\% to even as bad as 2\% of the continuum in the line core.

Telluric corrections in the short-K band provide additional challenges.
While there is no strong hydrogen line residing here, making early-A dwarfs
near perfect telluric tracers between 2 and 2.15 $\mu$m, the Earth's
atmosphere becomes increasing problematic below 2.08 $\mu$m.  Here, very strong 
and quickly changing telluric absorption is found, predominately due to 
the ro-vibrational transitions of CO$_2$.  Just the {\sl start} of this
very strong absorption is illustrated in Figure 1. The absorption continues
to get deeper at shorter wavelengths not given in Figure 1. For a thorough 
discussion of the special problems posed in this spectral region, see 
Kenworthy \& Hanson (2004).  The S/N measures listed in Table 1 
represent the average through out most of the spectral range covered for the
spectrum, and does not consider that over the CO$_2$ region, the S/N
can drop by 30\% or more.  Similar telluric noise can also be seen in the $H-$band 
region, where telluric features become increasingly strong and problematic 
longward of 1.72 $\mu$m (Fig.\ 1).

\section{THE SPECTRA}

This atlas contains $H$- and $K$-band spectra for 37 OB stars, ranging from O3 to
B3, and sampling most luminosity classes over that spectral range.  The 
spectra have been displayed in two ways.  All spectra in the atlas are
presented in Figures 2 through 8.  Here the spectra have been arranged based
on temperature so the reader can see the temperature-dependent
variations.  In Figures 9 through 12, we have presented the spectra grouped
by similar temperature, but arranged along varying luminosity. Here the 
reader can see the luminosity-dependent variations for four differing 
temperature regimes.

\subsection{Line Identifications}

The NIR is home to the Brackett Hydrogen series, beginning with
Brackett $\alpha$ at 4.052~$\mu$m (all wavelengths listed are given for
air).  Over the spectral range covered in this
atlas, we see $2.1661~\mu$m (4-7) Brackett $\gamma$ (Br$\gamma$), 1.736~$\mu$m 
(4-10) Br10, $1.681~\mu$m (4-11) Br11 and $1.641~\mu$m (4-12) Br12.

Lines of HeI and HeII both exist in the $H$-
and the $K$-band regions.  Within the $H$-band, one finds absorption due to
HeII $\lambda$1.693 (7-12) and HeI $\lambda$1.700 (3$p\ ^3$P$^o-4d\ ^3$D, triplet).  
Within the $K$-band, there exists the lines, HeII $\lambda$2.188 (7-10), 
HeI $\lambda$2.1127 ($3p\ ^3$P$^o-4s\ ^3$S, triplet) and $\lambda$2.1138
($3p\ ^1$P$^o-4s\ ^1$S, singlet). Though our spectral coverage from Subaru does
not include this line, still another important He line is the singlet
HeI transition, $\lambda$2.0581 ($2s\ ^1$S$-2p\ ^1$P$^o$).  This line is seen
in several of our late-O, early-B supergiants observed with ISAAC. 
The higher resolution and S/N obtained in this atlas
allows additional HeI lines to be identified.  In late-O and early-B supergiants
(Fig.\ 5, 7, 8, 11 \& 12) we see HeI $\lambda$2.161 ($4d\ ^3$D$-7f\ ^3$F, triplet) 
and at $\lambda$2.162 ($4d\ ^1$D$-7f\ ^1$F, singlet) begins 
to appear in the blue wing of Br$\gamma$. 

After Helium and Hydrogen, most OB stars show relatively few spectral features
in their NIR spectra.  Among the hottest stars, the triplet 
due to CIV ($3d^2D-3p^2P^o$) at 2.069, 2.078 and 2.083 is seen.  In our
O3 supergiant, Cyg OB2 \#7, the leading line in this set, at 2.078$\mu$m, 
is seen in absorption!  A second important metal line seen in the NIR spectra of
hot stars is the broad emission feature found at 2.1155$\mu$m, tentatively
identified by Hanson et al.\ (1996) as NIII (7-8), though may instead (or
also) be due to the very similar transition of CIII (7-8).  The difficulty 
in firmly identifying this features comes from the fact that the atoms share
 a similar structure and the transition identified here originates between 
very high lying levels.  It is expected, based on arguments of relative
abundances between N and C, that the feature seen is dominated by NIII,
particularly in more evolved stars. Stars where CIII may dominate the profile, 
would include hotter, less evolved stars which also exhibit strong CIV 
emission.  One can find additional discussion of the 2.1155$\mu$m feature in 
Repolust et al.\ (2005).

Additional lines previously cataloged in hot stars and seen in our spectra
include: $\lambda$2.100$\mu$m NV seen in HD 64568, an O3 V((f)), $\lambda 
2.137/2.143 \mu$m MgII in early-B supergiants, and numerous HeI lines at 
$\lambda 2.150 \mu$m, $\lambda 2.161/2 \mu$m, and $\lambda 2.184 \mu$m, all 
seen in late-O and early-B  supergiants (see Fig.\ 8).  Because of the high 
resolution and S/N of our spectra, a few new, unidentified lines 
have been found.  These include absorption lines seen in
mid and late-O stars at $\lambda 1.649, 1.651 \mu$m, an emission feature seen
in the O5 If+ star HD 14947 at $\lambda 2.1035 \mu$m, and several emission
features seen clustered around the Br10 line of the O3 If* star, Cyg OB2 \# 7.
The strongest of these, centered around $\lambda 1.735 \mu$m, might also be
seen in the O4 I(n)f star HD 66811.

\subsection{Classification Criteria for Temperature
and Luminosity}

As with the optical (Walborn \& Fitzpatrick 1990), the principal temperature 
classification criteria for O stars in the NIR is found within the behaviors
of the HeI and HeII lines. 
The strength of the He lines as a function of temperature has been well established
based on earlier, lower-resolution work (Hanson et al.\ 1996; Blum et al. 1997;
Hanson et al.\ 1998), which have quantified its behavior. Ionized Helium,
the strongest of the NIR transitions being the HeII (7-10) line at $\lambda$2.1885,
is present in absorption in essentially all O stars, regardless of luminosity class
(though it is so weak in O9 V stars it would go undetected in lower quality 
spectra). By mid-O, neutral Helium emerges (see Fig. 2, 4) and is retained until 
early-B for dwarf stars (see Fig.\ 3), but as late as B7 or B8 in supergiants 
(see Hanson et al.\ 1996).  When both HeII and HeI are present, temperature estimates 
are at their best.  

The unique HeI line at $\lambda$2.0581$\mu$m is highly sensitive to temperature
{\sl and} wind properties.  It first appears in absorption in mid- to late-O
stars (see Fig.\ 5, 7), and is frequently seen in strong emission in early-B 
supergiants (see Fig.\ 8).  Unfortunately, this particular study is unable to shed 
significant light on its behavior due to the small number of stars observed here.
It should be noted that this line lies within a region of the Earth's atmosphere 
where interfering telluric absorption is large, and accurate, high S/N profiles 
are challenging to obtain from the ground.

Unlike the Helium transitions which are unique to hot stars, hydrogen lines 
are instead ubiquitous to all stars of cosmic abundance. Despite being common, the 
hydrogen lines serve an important role in constraining characteristics of OB stars.  
Over the temperature range of this study, the cores of the Brackett transitions,
particularly Br10 and Br11, are excellent indicators of gravity.  Among stars
of similar rotational velocity, stars with lower gravity (higher luminosity),
are seen to have much deeper cores.  This leads to a {\sl stronger}, larger equivalent
width in the Brackett lines in high luminosity O and early-B stars (see Figs.\ 9, 10, 11 and 12).  
While some of the HeI lines appear to follow a similar route, their profiles 
becoming deeper and sharper as luminosity increases in OB stars, much of this
profile change is due to a change in rotational velocity (see for example,
Fig.\ 11). This behavior of deepening of the cores in the upper level Hydrogen
lines, is entirely explained theoretically.  Repolust et al.\ 
(2005) explain that the core is simply responding to the Stark-profiles which
are a strong function of electron density.  Such an effect is only seen in the cores
of hydrogen lines with large upper principal quantum numbers.  For Br$\gamma$,
with upper quantum number n=7, the effect is less sensitive and it is entirely 
insensitive in H$\alpha$ where the cores are instead dominated by Doppler-broadening.  
Thus the profile shape of the cores in Br10 and Br 11 provide a sensitive
indicator of gravity (luminosity), provided broadening from rotation is already
constrained \footnote{Unfortunately, a higher luminosity implies a larger mass-loss
  rate, so that the profiles might become refilled by wind-emission and
  thus weaker again. One such example is given in Fig. 9, if one
  compares Br12/11/10 from Cyg OB 8c (O5\,If) with those from HD\,14947
  (O5\,If+).}.  In high gravity OB stars, there is greater absorption in the wings
of the higher order Brackett profiles (see in particular Fig.\ 12). It is 
important to recognize that without adequate resolution these subtle changes 
in Brackett $\gamma$ and Helium line profiles which trace luminosity would 
be lost.

The metal lines of CIV and NIII can be used as vague temperature indicators
and are particularly useful with low S/N spectra or in spectra
plagued with very strong nebular emission lines rendering the helium
lines unusable.  CIV appears in emission starting around O4-O5,
and is seen down to about O7 in dwarf stars (though it is weak
and may be missed with low quality spectra).
This atlas lacks adequate sampling of the O8 spectral sequence in
luminosity.  However, the existence of CIV in O8 supergiants, and its increasing
strength with increasing luminosity, was shown in Hanson et al.\ (1996).  
NIII is seen in all early-O stars, down to about O7 in the dwarfs, and 
possibly as late as O8 in supergiants, though it is heavily blended with the
HeI $\lambda$2.112/3 line. In very luminous late-O stars,  the
triplet $\lambda$2.1127 ($3p\ ^3$P$^o-4s\ ^3$S) remains in absorption,
while the singlet $\lambda$2.1138 ($3p\ ^1$P$^o-4s\ ^1$S) line goes 
into emission (Najarro et al.\ 1994).  As previously discussed,
a confident identification of the line at 2.1155~$\mu$m (whether it be CIII or 
NIII) is still lacking.   For purposes of spectral classification, this is not an issue.   
For purposes of spectral analysis, it will be.

\section{DISCUSSION}

In the mid-90s, several papers presented low and moderate resolution 
NIR spectra of stars.  Previous to spectroscopic studies, the only 
tool for determining the characteristics of heavily reddened stars was 
through broadband NIR colors.  While such measures are effective for
constraining cool stars, NIR photometric colors becoming increasingly
degenerate for hot stars.  It was pointed out by Massey et al.\
(1995) that UBV colors were degenerate for O stars, necessitating 
the need for MK classification for proper determinations of mass
functions within OB clusters.  For JHK colors,
the degeneracy begins at A stars.  The development of even a low
resolution classification system for the NIR has proved 
enormously useful for those studying hot stars behind significant 
interstellar extinction.

However, with near-infrared spectrometers becoming more sophisticated, it was
time to push the observations {\sl and theory} to a complementary level.
Our groups theoretical work shows that with increased resolution 
and S/N rather accurate physical characteristics of
hot stars can be derived with NIR spectra alone (Repolust et al.\
2005).  But this does require an analysis be used in conjunction 
with theoretically derived line profiles to achieve such accurate
characteristics.

\subsection{What level of S/N and resolution will be required for a
quantitative analysis?}

It's unlikely that typical researchers will have the luxury that this
study possessed: using an 8-m class telescope to observe exceedingly 
bright, single sources, void of nebular contamination.  
In many astrophysical situations, nebular emission can 
contaminate important diagnostic lines (most usually
Br$\gamma$ and $\lambda 2.058\mu$m HeI), rendering them useless. 
In the most extreme situations, where the extinction to the massive
star is exasperated by thermal emission in the NIR from
nebular or circumstellar material, all 
absorption lines may be undetectable. It is clear that 
both NIR spectral 
analysis and classification will have very real 
and frustrating limits in their application 
to massive young stellar objects (Hanson et al.\ 1997; 
Blum et al.\ 2001; Bik et al.\ 2005a).  
In incidences where there is significant contamination from thermal
emission, this has an effect much like reducing the S/N.  Imagine if one
obtained a stellar spectrum with a S/N of 200, but half of the photons originate
in a featureless continuum generated by a disk.  The depth and strength 
of the stellar  
lines would be reduced by about one half.  This would be roughly equivalent 
to obtaining the spectrum of an undiluted star, 
but with half the S/N.  If it is believed that
continuum contamination is roughly equal to the stellar flux, then
improving S/N might allow the detection of the stellar lines. 
Spectroscopic systems designed to be used with adaptive optics 
would reduce the contamination from extended 
emission and increase the likelihood of detecting stellar features.
In cases where there may be significant thermal contamination,
classification should not be based on equivalent width measures 
alone, but by the relative behavior of critical line pairs (see \S 5.2).

Ignoring for a moment contamination from nebular or circumstellar
emission, what advice might we give to those interested in doing
a more accurate spectral analysis in the NIR?  Unfortunately, the
resolution and S/N limits required are a function of the stars
spectral type, luminosity class and rotational velocity (something 
the researcher does not know at the start!) and whether one
wishes to use a simple classification scheme or a full blown
profile analysis.  This is realized 
when one looks closely at the line strengths among the O dwarfs.
In Figure 13, we plot the central region of the $H-$band, which 
happens to contain a Hydrogen line as well as neutral and ionized
Helium.  To create this figure, we created a set of low S/N line-free 
spectral regions from our data.  These have been multiplied against 
are true spectra to illustrate the effect of reduced S/N on the detection
of these weak stellar features.  All the spectral features shown 
are $<$ 2~\AA\ in
strength, the strongest line being the HeI line at 1.70 $\mu$m in
the O9 V star HD~149757 (e.w. = 1.8\AA).  Once the S/N was reduced to
150, the weak (0.45 \AA) HeII feature at 1.693 $\mu$m in HD 217086, 
is no longer confidently detected.  For nearly all early O stars, 
we suggest a S/N $>$ 150 for a quantitative analysis, to detect the 
very weak features and properly match the wings in the line.
For late-O and B stars, such as in HD 149757, keeping S/N $>$ 100 
should be sufficient for their slightly stronger lines.  Crude
classification can still be obtained with a S/N under 100, but
just barely.  Only the strongest lines, EW $>$ 2.0~\AA\, can 
be confidently detected once the S/N drops to 50 (Fig.\ 13).

For a proper profile analysis (and to resolve blends important in
diagnosing luminosity classes) a resolution of at least $\lambda 
/ \delta \lambda > 5000$ should be adhered to. Such a resolution is
well matched for many if not most OB stars, which typically possess
fairly high rotational rates ($V {\rm sin} i > 100$ km/s). However, 
for OB stars with low rotation ($V {\rm sin} i < 100$ km/s), 
a resolution of perhaps 8,000 or more will be needed to resolve their
underlying profile (see, for instance, the slow rotator, HD 149438,
Tau Sco, in Fig.\ 3).  While seemingly high (by NIR standards!), 
these resolutions are really very low.  Similar studies on cooler 
stars require much higher resolutions of several to many tens of 
thousands for a proper analysis (e.g., Luck \& Heiter 2005).  

\subsection{The use of equivalent width for classification}

In nearly all previous NIR spectral atlases, enormous tables and
figures are given showing the equivalent width  of strategic lines as a 
function of spectral type or luminosity class. Such measures have been 
critical to the development of a classification tool for the NIR.
However, we do not present such tables here.  The spectra in
this atlas differ greatly from previous atlases, and thus
dictates a different mind set for their use.  We suggest that when
working with high S/N and moderate to high resolution spectra, 
even without the use of quantitative profile analysis, one must
adopt the philosophy outlined decades ago for the classification
of stars in the optical.  Here classification is based on {\sl the
comparison of spectra}; that between stars of known spectral and 
luminosity class and stars which are unknown.  Important
for a proper comparison, the resolution and approximate
S/N of the stars being compared must be the same.   What has been 
sorrily missing, 
however, in most NIR spectral classification programs 
has been {\bf spectroscopic standards}.  There isn't
a single, conscientious optical spectroscopist doing 
classification work that would dream of skipping the step
of taking classification standards.  This is for a 
field for which optical classification spectra must number
in the millions.  In the infrared, researchers are relying on
the scant, 100 or so O stars observed to date, to make
crude comparisons to their own, independently obtained 
NIR spectra.  This level of crude judgment to derive temperature
and luminosity should not be tolerated if NIR classification is 
to become a robust technique. Moreover, as mentioned in \S5.1,
thermal contamination, something most optical astronomers do not
typically deal with, makes classification via equivalent width
strengths a very risky method for some applications.

The importance of obtaining ones own set of classification
standards was recently highlighted in a study by Bik et al.\
(2005b).  They obtained high S/N, moderate resolution NIR
spectra of ionizing stars of IRAS selected young star forming regions.
Within the realm of this study, the researchers obtained 
spectra for a number of known O and early-B dwarf stars, using
the identical spectroscopic set up and typical S/N obtained for 
their heavily reddened ionizing sources.  They found a small but
significant difference in the line strengths of
their standard stars compared to previous lower resolution,
lower S/N NIR atlases.  Without the ability
to calibrate with their own set of classification standards,
they would have miss-classified most of their sources.

Looking again to the optical for guidance, most classifications 
criteria are based on the comparison of two lines in 
a spectrum.  Typically, sub-classes are defined by when
a set of lines are of equal strength, one line is stronger
than the other, or when a line first appears or disappears, 
etc.\ (Jaschek \& Jaschek 1987).  
While at first this method may appear crude, 
it in fact makes for a highly repeatable 
(from person to person) evaluation and is the crux to the 
success of the optical classification system.   
In the NIR we do not have such a wealth of diagnostic lines to
allow for comparisons between oppositely behaving
spectral features over all temperature and luminosity
ranges of interest.  The He lines do offer such a system
to constrain temperature once O stars get cool 
enough to show HeI, and before they get too cool to show HeII
(over the very narrow range of O7 to about O9.5).  However,
when including comparisons with other lines, CIV, Br$\gamma$,
reasonable estimations appear possible.  We have tried
to outline just those comparisons within this paper (\S 4.2),
though we admit, the field of NIR spectral classification is
still very young. Still more observations are need to establish 
``typical'' spectroscopic behavior in the NIR.

\section{CONCLUSIONS}

We present intermediate resolution (R $\sim$ 8,000 - 12,000) high 
S/N $H$- and $K$-band spectroscopy of a sample of optically visible O and early-B stars.  
The purpose of this study is to better characterize OB spectral profiles 
in the NIR.  We have also established some observational limits for researchers
preparing to use a quantitative analysis to derive stellar
temperature and luminosity with NIR spectra alone.
In order to directly determine effective
temperatures and log$g$ for individual stars, one needs to work 
with atmospheric model codes which rely on profile fits.  Such 
programs have recently been developed and show great promise 
(Lenorzer et al.\ 2004; Repolust et al.\ 2005). For most
stars, a S/N $> 100$ and resolution of $r \approx 5000$ should be 
just sufficient.  However, if the targets of interest turn out to
be early-O, or have a very low rotational velocity, a higher
S/N ($> 150$) or resolution (R $> 8000$), respectively, will need to 
be obtained.  

When NIR spectral classification is sought, we {\sl strongly 
encourage researchers to obtain spectra of known OB stars} 
for direct comparisons to their target stars.  Until a time
when equivalent ``MK'' standards are developed in the NIR, optically 
studied and thus known to be 'well-behaved' stars which bracket 
the expected spectral and luminosity range of the targets should 
be sufficient for this purpose.

\acknowledgements

We gratefully acknowledge the Subaru and VLT Observatories
for their support of our program.  This research has made use of the NASA's Astrophysics
Data System Bibliographic Services and the SIMBAD database operated at
CDS (Strasbourg, France). MMH and MAK gratefully acknowledge support
for this program from the National Science Foundation under 
Grant AST-0094050 to the University of Cincinnati.

{\it Facilities:} \facility{VLT:Antu (ISAAC)}, \facility{Subaru (IRCS)}

\appendix

\section{Appendix}

While not part of the original science goals of this program, the extraordinary 
spectral resolution and signal-to-noise achieved on all stars in this
study, including the standard stars, seemed too good to go unpublished.
Both for illustrative purposes, as well as for use by those
looking to model and remove Brackett series transitions from their
A-dwarf telluric standard stars, we provide spectra for all
the A dwarf stars observed as part of the Subaru-IRCS program. Table A.1
gives the stars and spectral types, Figure 14 shows the spectra. 
Because we intend for these spectra to be used as templates, we have
removed most of the high order noise features artificially.  This was
not done in the OB spectra because we did not want to inadvertently 
alter their very weak profiles.  

The stars in Figure 14 have been arranged first by spectral class (all are dwarfs), and
next by rotational velocity.  Over even this small sample, its clear that
in modeling (to remove) Brackett features in A-dwarf stars, knowing the
rotational velocity is at least as important as knowing  the 
spectral type. The profile differences are small 
compared to the differences seen as a function of rotational velocity.

\begin{deluxetable}{lcccccc}
\tablewidth{0pt}
\tablecaption{Sample and Observing Data}
\tablehead{
\colhead{Star} &  
\colhead{SpType} &
\colhead{2MASS $H,K$} &
\colhead{$\alpha$(2000)} &
\colhead{$\delta$(2000)} &
\colhead{VLT-ISAAC$^a$} &
\colhead{Subaru-IRCS$^b$} 
}
\startdata 
Cyg OB2 \#7 & O3 If$^{\ast}$ & 6.8, 6.6 & 20 33 14.1  & +41 20 22 &  ...  &  Nov 01   \\
Cyg OB2 \#8A & O5.5 I(f) &  5.7, 5.5 &20 33 15.1 & +41 18 50 & ... & July 02   \\
Cyg OB2 \#8C & O5 If &  6.8, 6.5 &20 33 18.1  &  +41 18 31 & ... & July 02  \\
\object{HD 5689}   &  O6 V   &  8.3, 8.4 & 00 59 47.6 & +63 36 28  &  ...  &  Nov 01/July 02   \\
\object{HD 13268}  &  ON8 V  & 7.9, 7.9 & 02 11 29.7  &  +56 09 32  &  ...  &  Nov 01 \\
\object{HD 13854} & B1 Iab   & 5.7, 5.6 &  02 16 51.7 & +57 03 19  & ... & Nov 01  \\
\object{HD 13866}  &  B2 Ib  & 7.1, 7.1 & 02 16 57.6 & +56 43 08 &  ...  &  July 02  \\
\object{HD 14134}  &  B3 Ia  & 5.4, 5.3 & 02 19 04.5 & +57 08 08 &  ...  &  July 02  \\
\object{HD 14947} & O5 If+   & 6.9, 6.9 & 02 26 47.0 & +58 52 33  &  ...  & Nov 01  \\
\object{HD 15570} &  O4 If+  & 6.3, 6.2 & 02 32 49.4 & +61 22 42  &  ...  & Nov 01  \\
\object{HD 15558}  & O5 III(f)  & 6.6, 6.5 & 02 32 42.5 & +61 27 22  &  ... & July 02  \\
\object{HD 15629}  & O5 V((f))  & 7.3, 7.3 & 02 33 20.6 & +61 31 18 &  ... & July 02   \\
\object{HD 30614}  &  O9 Ia  & 4.4, 4.2 & 04 54 03.0 & +66 20 34  &  ...  &  Nov 01   \\
\object{HD 36166}  & B2 V    & 6.3, 6.3 & 05 29 54.8 & +01 47 21  &  ...  &  Nov 01   \\
\object{HD 37128}  & B0 Ia   & 2.4, 2.3 & 05 36 12.8 & $-$01 12 07  &  ...  &  Nov 01   \\
\object{HD 37468}  & O9.5 V  & 4.6, 4.5 & 05 38 44.8 & $-$02 36 00  &  ...  &  Nov 01   \\
\object{HD 46150}  & O5 V((f)) & 6.5, 6.4 &  06 31 55.5 & +04 56 34 & ...  &  Nov 01   \\
\object{HD 46223}  & O4 V((f)) & 6.7, 6.7 & 06 32 09.3 & +04 49 24 & ...  &  Nov 01   \\
\object{HD 64568}  &  O3 V((f)) & 9.1, 9.1 & 07 53 38.2 & $-$26 14 03 & ... & Nov 01  \\
\object{HD 66811}  &  O4 I(n)f & 3.0, 3.0 & 08 03 35.0 & $-$40 00 11  & ... & Nov 01  \\
\object{HD 90087}  & O9.5 II  & 7.7, 7.8 & 10 22 20.9 & $-$59 45 20  &  May 01    & ...   \\
\object{HD 92554}  & O9 IIn   & 9.1, 9.1 & 10 39 45.7 & $-$60 54 40  &  May 01    & ...   \\
\object{HD 115842} & B0.5 Ia & 5.3, 5.2 & 13 20 48.3  & $-$55 48 03 & Apr 01 & ...  \\
\object{HD 123008} & ON9.5 Iab & 7.8, 7.7 & 14 07 30.6 & $-$64 28 09 &  Apr 01 & ...  \\
\object{HD 134959} & B2 Ia   & 5.5, 5.2 & 15 15 24.1  & $-$59 04 29 &  Apr 01 & ...  \\
\object{HD 148688} & B1 Ia   & 4.1, 4.1 & 16 31 41.8 & $-$41 49 02 &  May 01 & ...  \\
\object{HD 149438} & B0.2 V  & 3.5, 3.7 & 16 35 53.0 & $-$28 12 58 & ... &  July 02  \\
\object{HD 149757} & O9.5 V  & 2.7, 2.7 & 16 37 09.5 & $-$10 34 02 &  ... & July 02  \\
\object{HD 154368} & O9.5 Iab & 4.9, 4.8 & 17 06 28.4 & $-$35 27 04  &  Apr 01  & ...  \\
\object{HD 163181} & BN0.5 Ia & 5.1, 4.9 & 17 56 16.1 & $-$32 28 30  &  Apr 01  & ...   \\
\object{HD 190864} & O6.5 III & 7.3, 7.3 & 20 05 39.8 & +35 36 28 & ...  &  July 02   \\
\object{HD 191423} & O9 III  & 7.7, 7.8 & 20 08 07.1 & +42 36 22 & ...  & July 02   \\
\object{HD 192639} & O7 Ib   & 6.3, 6.2 & 20 14 30.4 & +37 21 13 & ...  & July 02   \\
\object{HD 203064} & O7.5 III & 5.1, 5.1 & 21 18 27.2 & +43 56 45 & ...  & July 02   \\
\object{HD 209975} & O9.5 Ib  & 4.9, 4.9 & 22 05 08.8 & +62 16 47 & ...  & July 02   \\
\object{HD 210809} & O9 Iab  & 7.4, 7.4 & 22 11 38.6  & +52 25 48 & ...  & July 02    \\
\object{HD 217086} & O7 V    & 6.1, 6.0 & 22 56 47.2 & +62 43 38  &  ...  &  Nov 01   \\
\enddata
\tablenotetext{a}{The VLT-ISAAC $H-$band resolution is R $\sim$ 10,000; the 
$K-$band resolution is R $\sim$ 8000. The typical S/N ratios obtained with 
these spectra were S/N $\sim$ 100-150, with areas as high as S/N $\sim$ 200, 
and as low as S/N $\sim$ 50, depending on the local telluric contamination}
\tablenotetext{b}{The Subaru-IRCS $H-$ and $K-$band band resolution is 
R $\sim$ 12,000.  The typical S/N ratios obtained with these spectra were 
S/N $\sim$ 200-300, with areas as high as S/N $\sim$ 500, and as low as 
S/N $\sim$ 100, depending on the local telluric contamination}
\end{deluxetable}

\clearpage

\begin{deluxetable}{lcccc}
\tabletypesize{\footnotesize}
\tablecaption{The OB Star Sample}
\tablehead{      &      V     &    III    &    II/Ib    &   Iab/Ia }
\startdata

O3               &  HD 64568  &            &         &  Cyg OB2 \#7  \\
\\
O4               &  HD 46223  &            &         & HD 66811, HD 15570 \\
\\
O5               & HD 46150, HD 15629 &   HD 15558 ($K$-band only) &   &   HD 14947, Cyg OB2 \#8C \\
                 &              &           &          &  Cyg OB2 \#8A \\      
O6               &  HD 5689   &           &           &              \\
                 &            &  HD 190864 \\
O7               & HD 217086  &            &   HD 192639         \\
                 &             &  HD 203064  \\
O8               & HD 13268 \\
\\
O9               &           &  HD 191423   & HD 92554   & HD 30614, HD 210809 \\
               & HD 37468, HD 149757 &      & HD 90087, HD 209975 &  HD 123008, HD 154368  \\
B0               &            &             &           & HD 37128  \\
                 & HD 149438  ($K$-band only)     &          &         & HD 115842, HD 163181   \\
B1               &            &             &          & HD 13854, HD 148688  \\
           \\
B2               &  HD 36166    &             &  HD 13866  &  HD 134959  \\
\\
B3               &            &             &           & HD 14134  \\
\enddata
\end{deluxetable}

\clearpage

\begin{deluxetable}{lcc}
\tablewidth{0pt}
\tablenum{A.1}
\tablecaption{Telluric Stars}
\tablehead{
\colhead{Star} &  
\colhead{SpType} &
\colhead{Vsin{\it i} (km/s)$^a$}
}
\startdata  %
HD 12279  & A1 Vn  & 275   \\
HD 28780  & A1 V   & 28   \\
HD 146624  & A0 V  &  39    \\
HD 171149  & A0 V  &  301  \\
HD 199629  & A1 V  &  217  \\
HD 205314  &  A0 V  &  191 \\ 
\enddata
\tablenotetext{a}{From Royer et al.\ (2002a, 2002b).}
\end{deluxetable}

\clearpage

\begin{figure}
\epsscale{0.8}
\plotone{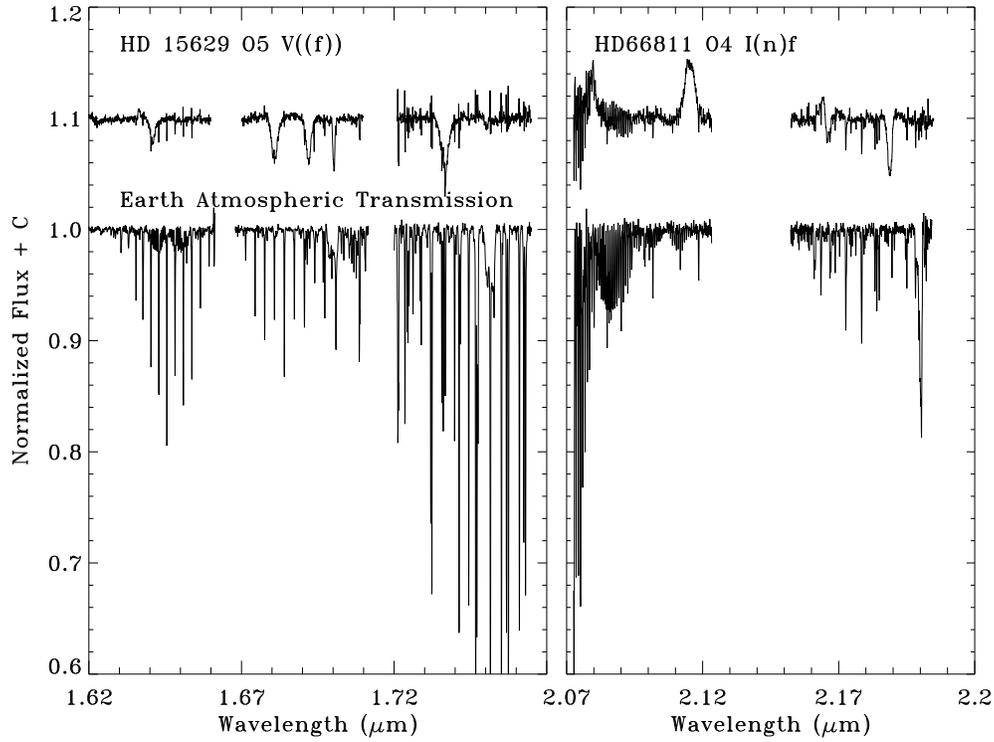}
\caption{{\bf Earth Telluric Lines.} The lower graph shows the wavelength 
dependence of the Earth's transmission, 
normalized to 1 to show only the absorption in small structure lines. On top
is shown the final reduced spectrum from two stars, HD15629 in the $H-$band and
HD 66811 in the $K-$band.  Telluric corrections can prove very challenging when 
there are either intermittent clouds (as was the case for HD 15629), or when the airmass is very
large and suitable standards sampling such airmass are not available (as was the case for
HD 66811).  The numerous noise spikes seen in the stellar spectra can easily be 
traced back to very strong narrow telluric features in the Earth's atmosphere. \label{fig1}}
\end{figure}

\clearpage

\begin{figure}
\epsscale{0.8}
\plotone{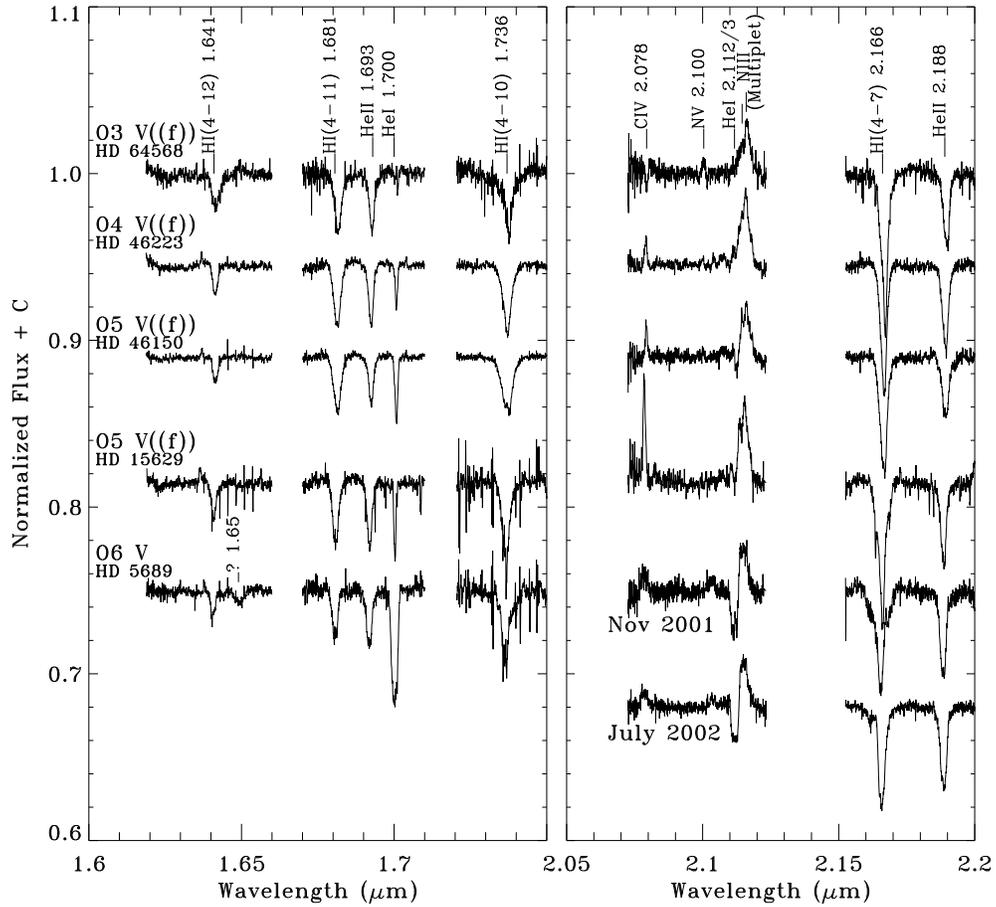}
\caption{{\bf Early-O Dwarf Stars.} Characteristic lines include strong
HeII ($\lambda 1.693\mu$m, 2.188$\mu$m), and among the hottest O3, no HeI nor
CIV emission (though possibly CIV absorption).  All early-O dwarf stars show
NIII emission at 2.1155$\mu$m. \label{fig2}}
\end{figure}

\clearpage

\begin{figure}
\epsscale{0.8}
\plotone{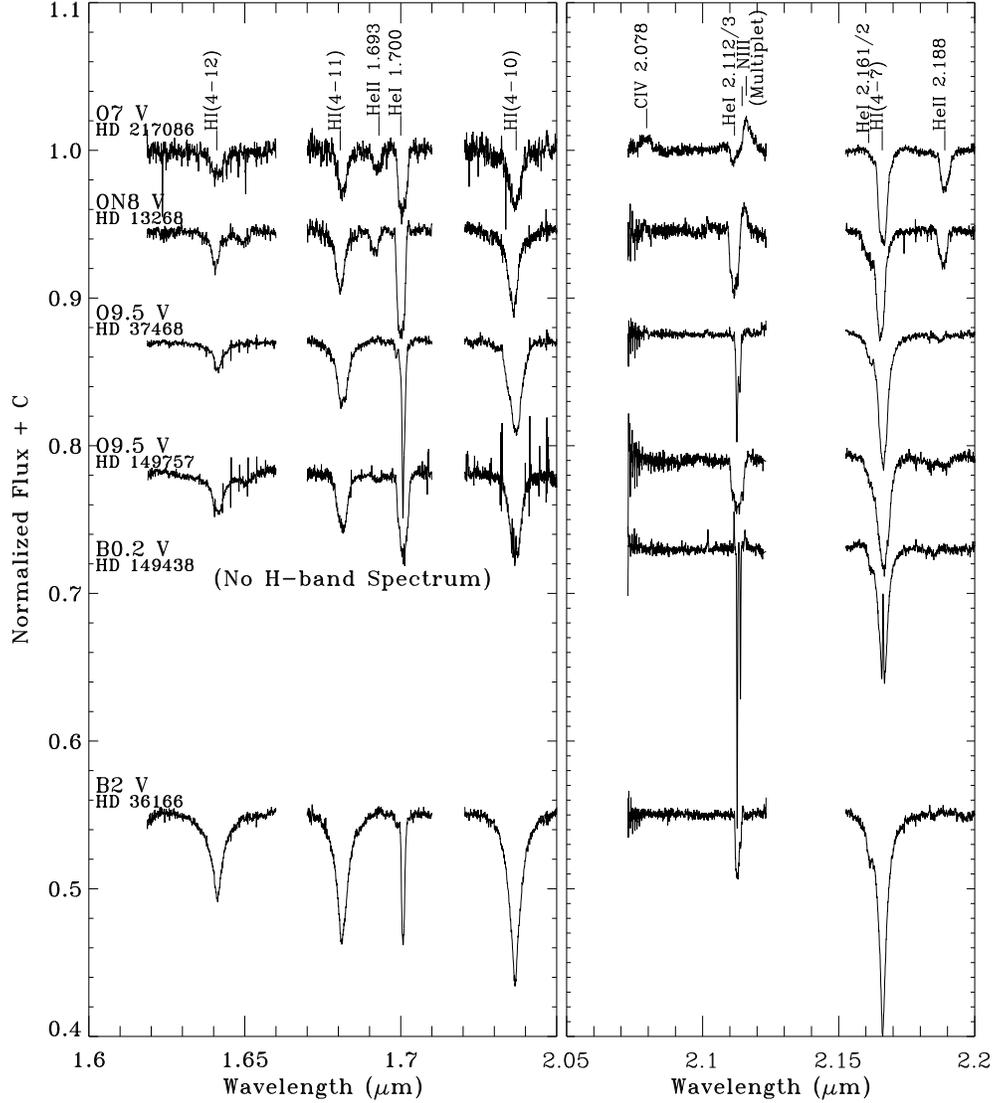}
\caption{{\bf Late-O Dwarf Stars.} Characteristic lines in late-O dwarfs include
simultaneously occurring HeI and HeII.  HeII disappears for B dwarfs
(just as in the optical 4686\AA).  We also see a rather strong HeI absorption
developing in cooler B dwarfs. Note the very deep and narrow HeI 
absorption in HD 149438, a very slowly
rotating star (Vsin$i \approx$ 10-20 km/s). \label{fig3}}
\end{figure}

\clearpage

\begin{figure}
\epsscale{0.8}
\plotone{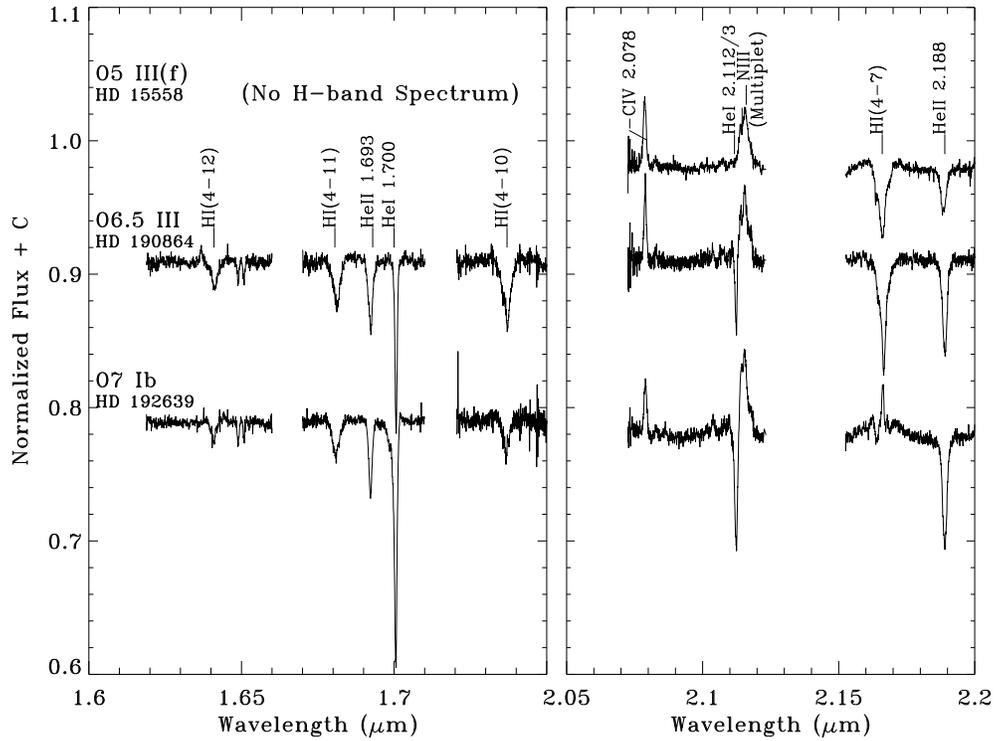}
\caption{{\bf Mid-O Giant Stars.} The effect of increased luminosity for this
spectral range is most directly seen in deeper Brackett and HeI absorption, 
though the line width is further narrowed by the slow rotation rate of these 
stars.  \label{fig4}}
\end{figure}

\clearpage

\begin{figure}
\epsscale{0.8}
\plotone{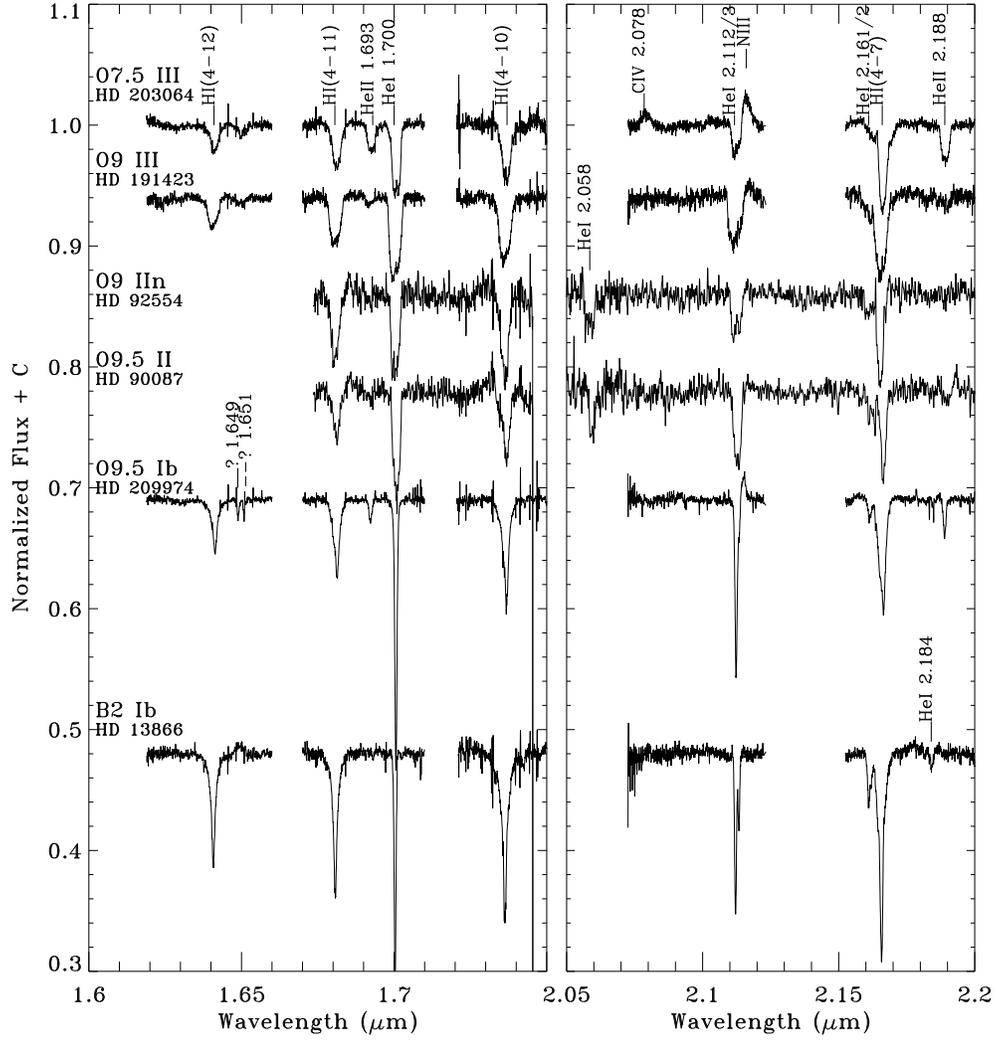}
\caption{{\bf Late-O and Early-B Giant Stars.} As shown in Fig.\ 3, these stars
show a deepening of their Brackett and HeI lines, as compared to dwarf stars of
the same spectral type.  HeI, at $\lambda 2.184\mu$m becomes apparent in early-B
giants. \label{fig5}}
\end{figure}

\clearpage

\begin{figure}
\epsscale{0.8}
\plotone{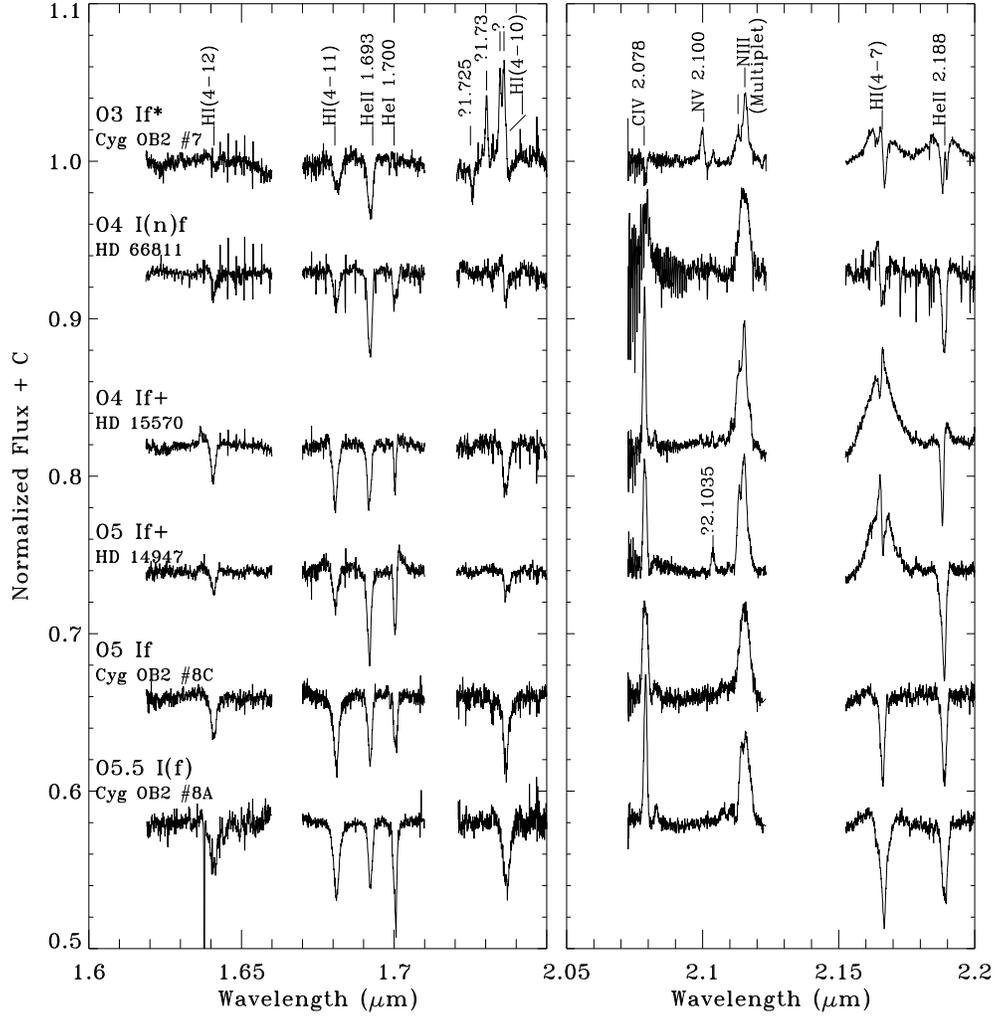}
\caption{{\bf Early-O Supergiant Stars.} In the most extreme early-O supergiants, Br$\gamma$
becomes partly or wholly filled in, or in strong broad emission.  Other signs of high
luminosity for this spectral class include, the narrowing and deepening of the Brackett and
Helium lines. \label{fig6} The unidentified doublet features marked as ``?'' in Cyg OB2 \#7 
were found to lie at $\lambda$1.7347,1.7360.}
\end{figure}

\clearpage

\begin{figure}
\epsscale{0.8}
\plotone{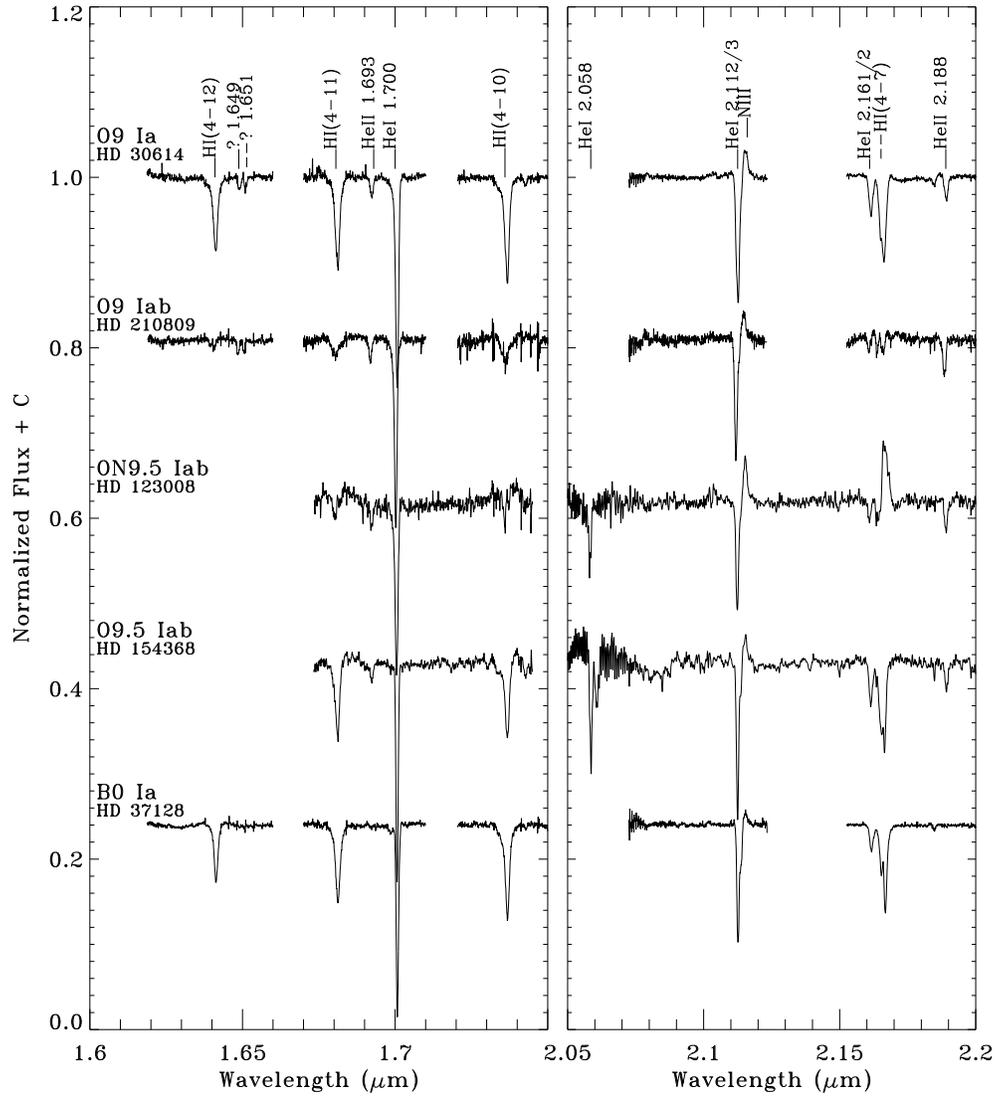}
\caption{{\bf Late-O Supergiant Stars.} Entirely unique to this range of 
temperature and spectral class is the appearance of weak, narrow HeI absorption 
at 2.161/2$\mu$m.  Both the HeI and Hydrogen Brackett lines have become extremely 
narrow and deep.  \label{fig7}}
\end{figure}

\clearpage

\begin{figure}
\epsscale{0.8}
\plotone{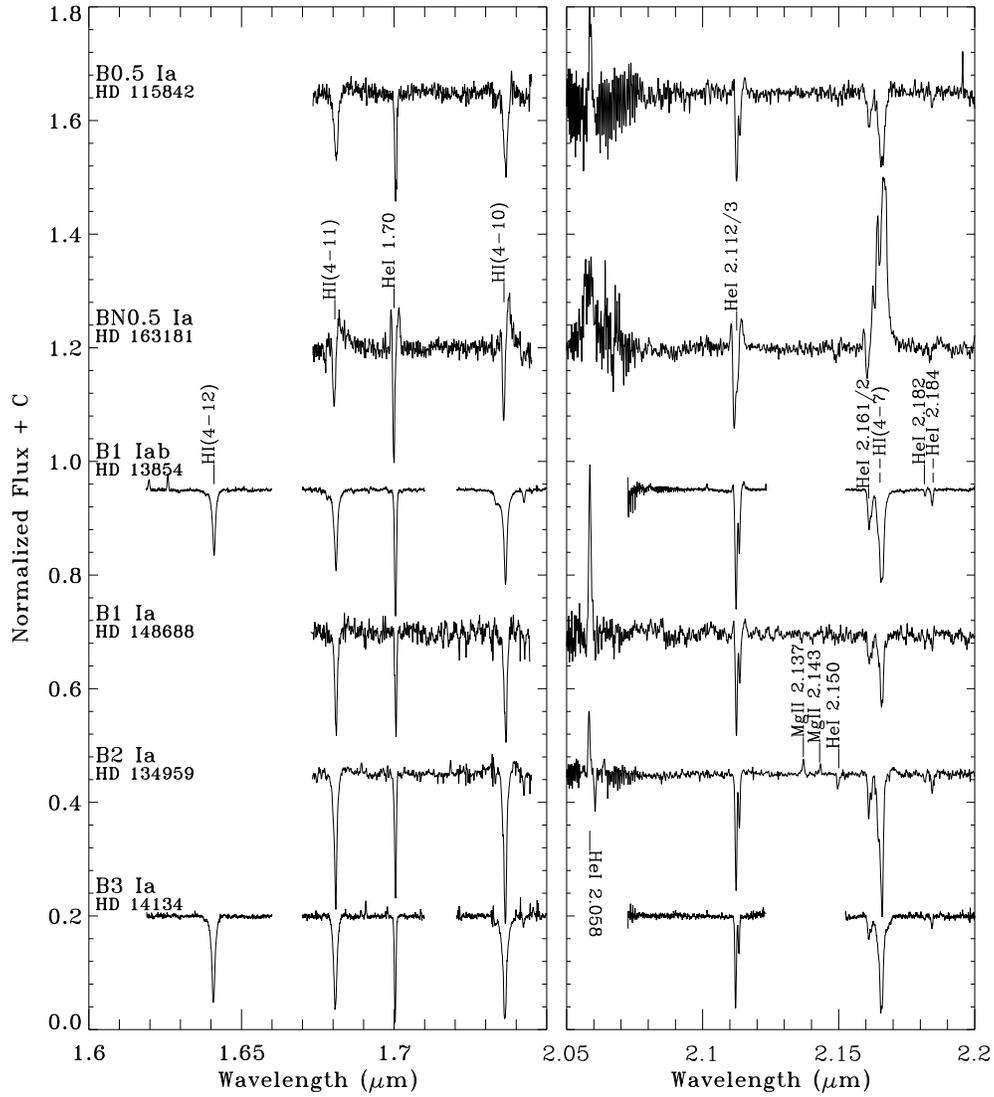}
\caption{{\bf Early-B Supergiant Stars.} As in Fig.\ 6, the HeI and Brackett lines
become deep and narrow.  However, distinct from their slightly hotter cousins,
the early-B supergiants also show weak, but significant HeI absorption at
2.184$\mu$m. \label{fig8}}
\end{figure}

\clearpage

\begin{figure}
\epsscale{0.8}
\plotone{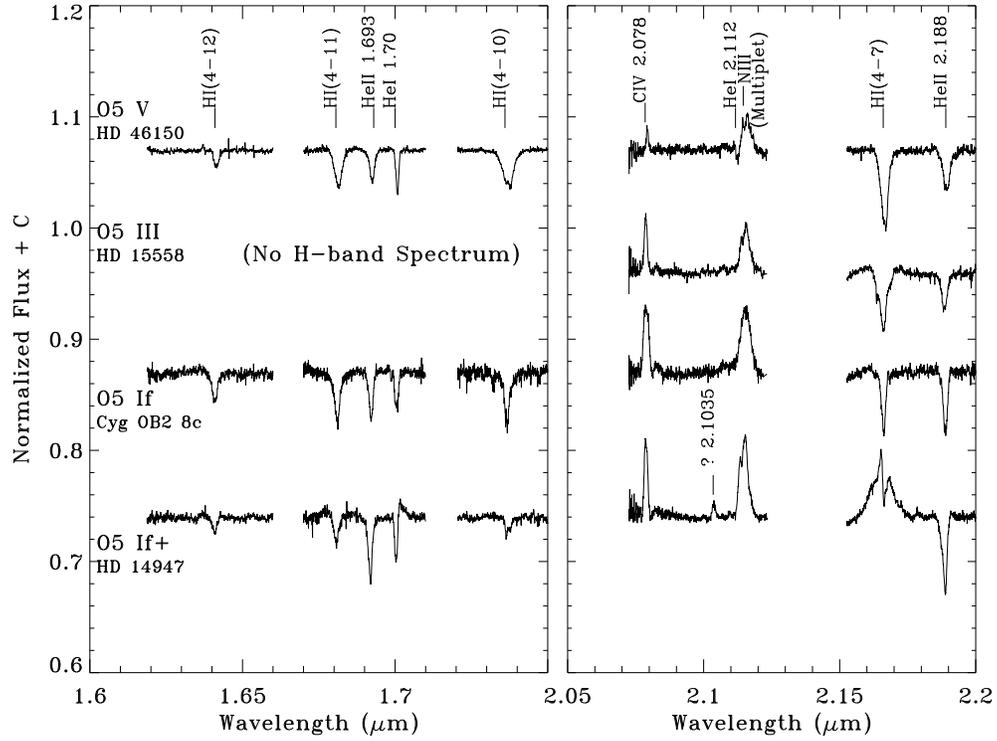}
\caption{{\bf Luminosity variations seen in O5 stars.} Deepening and narrowing of the
core in the Brackett line and loss of broad wings correlates with increased luminosity
in O5 stars.
In the most extreme O supergiants, Br$\gamma$ is in emission.  
Note the contamination of Br12/11/10 in HD14947 by wind-effects (see
  text \S 4.2).\label{fig9}}
\end{figure}

\clearpage

\begin{figure}
\epsscale{0.8}
\plotone{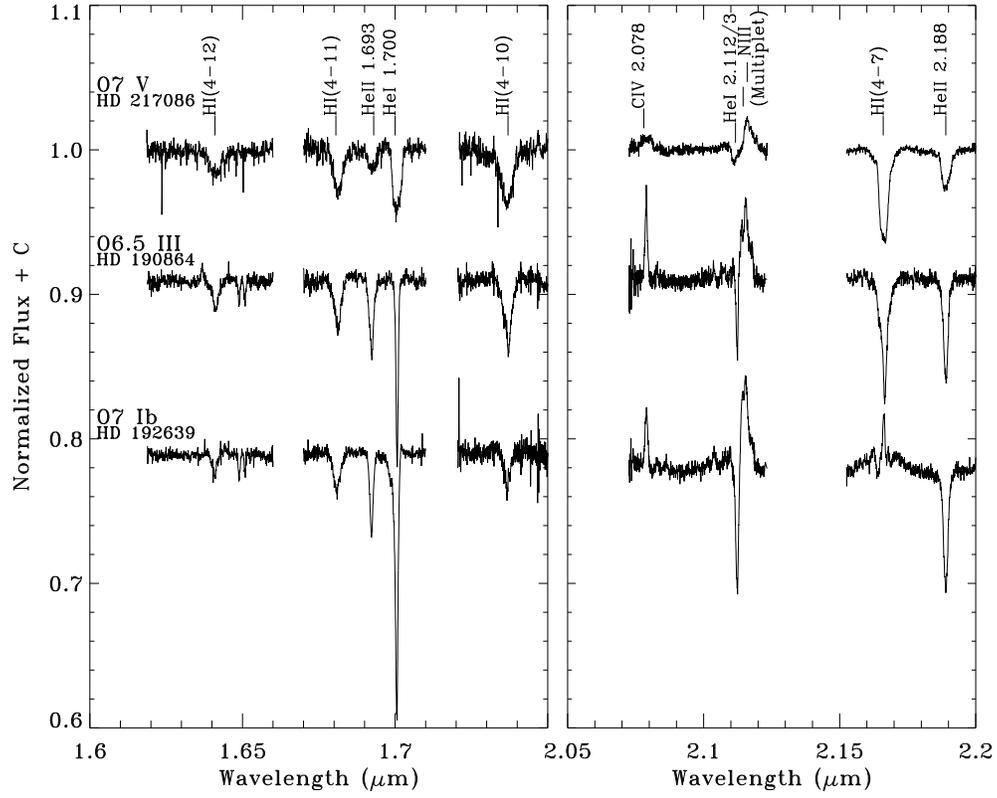}
\caption{{\bf Luminosity variations seen in O7 stars.} The same effects are seen in the O7 stars
as was displayed in the O5 stars, in Fig.\ 9.  \label{fig10}}
\end{figure}

\clearpage

\begin{figure}
\epsscale{0.8}
\plotone{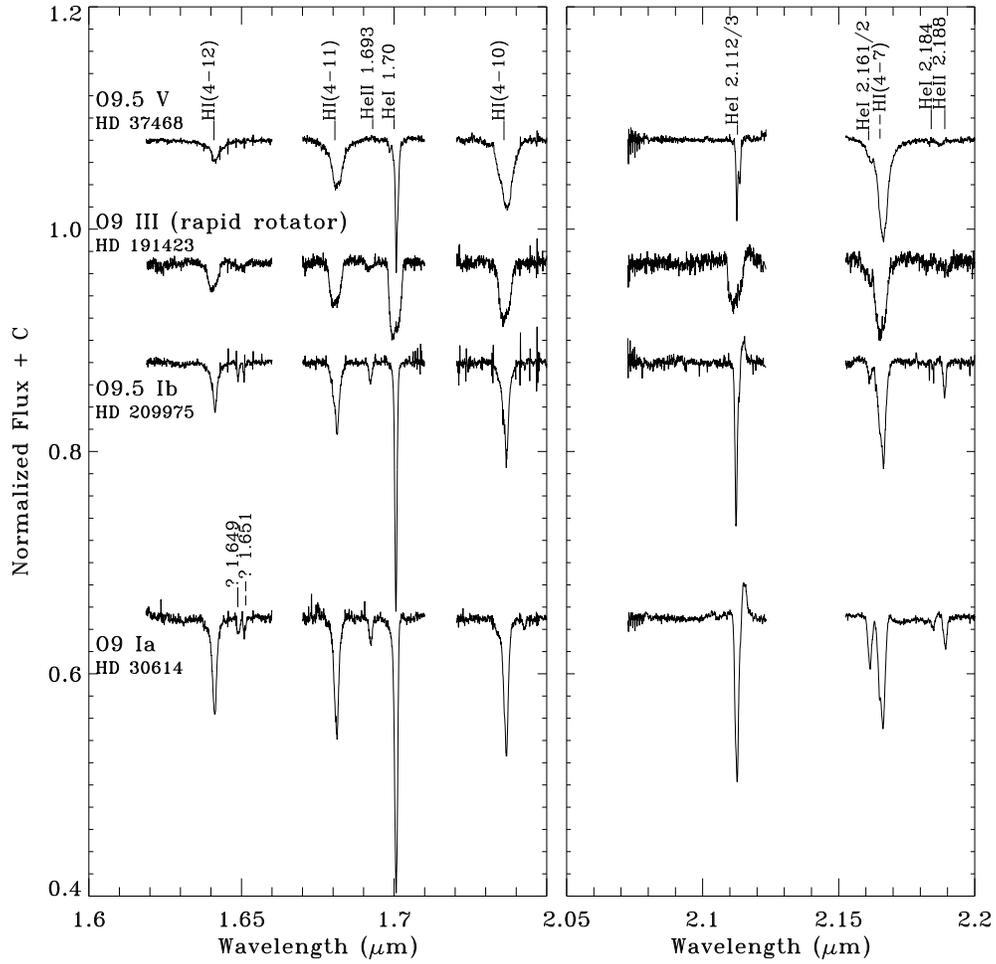}
\caption{{\bf Luminosity variations seen in O9 stars.} It is interesting to note that the O9 giant
might be incorrectly classified as a dwarf, if it wasn't known to be a rapid rotator.  The broad
HeI at 1.700$\mu$m gives away its large Vsin$i$.  \label{fig11}}
\end{figure}

\clearpage

\begin{figure}
\epsscale{0.8}
\plotone{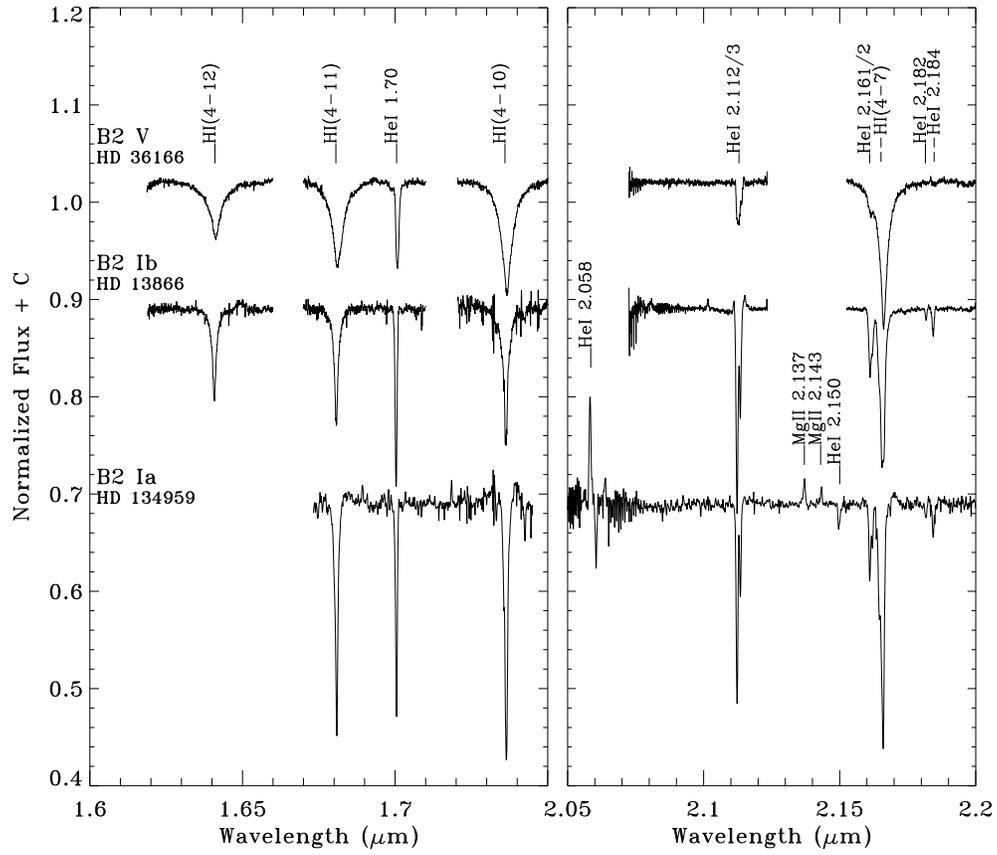}
\caption{{\bf Luminosity variations seen in B2 stars.} Lack of wings, deepening cores, presence of
HeI at 2.161/2, 2.184$\mu$m all correlate with increased luminosity among early-B stars.  \label{fig12}}
\end{figure}

\clearpage

\begin{figure}
\epsscale{0.8}
\plotone{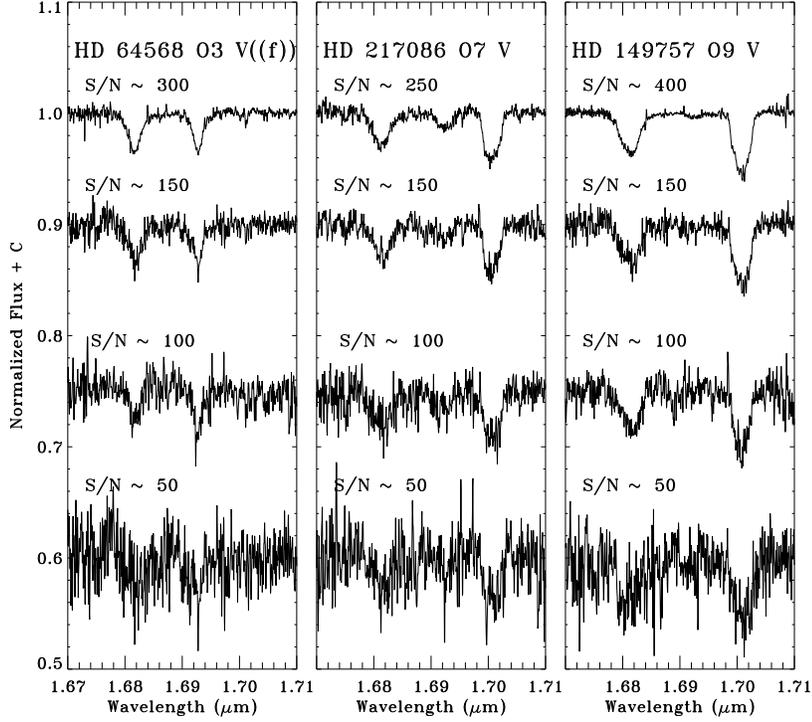}
\caption{{\bf The need for high signal-to-noise in OB star spectra.} This figure shows 
our final spectra plotted on the top of each panel. Below those spectra, the same 
spectra are given, but the S/N has been artificially reduced for illustrative purposes. 
Very high signal-to-noise, 
S/N $>$ 150, will typically be required for a reliable quantitative analysis performed on the
stellar profiles.  The strength of 
features presented range from e.w. = 0.42~\AA\ for the 1.693~$\mu$m HeII line in HD217068 to
e.w. = 1.8~\AA\ for the 1.700~$\mu$m HeI line in HD~149757.  If the signal-to-noise 
drops much below 100, even spectral classification becomes unreliable, as features with
strengths below $\sim 1.0$~\AA\ become undetectable. \label{fig 13}}
\end{figure}

\clearpage

\begin{figure}
\epsscale{0.8}
\plotone{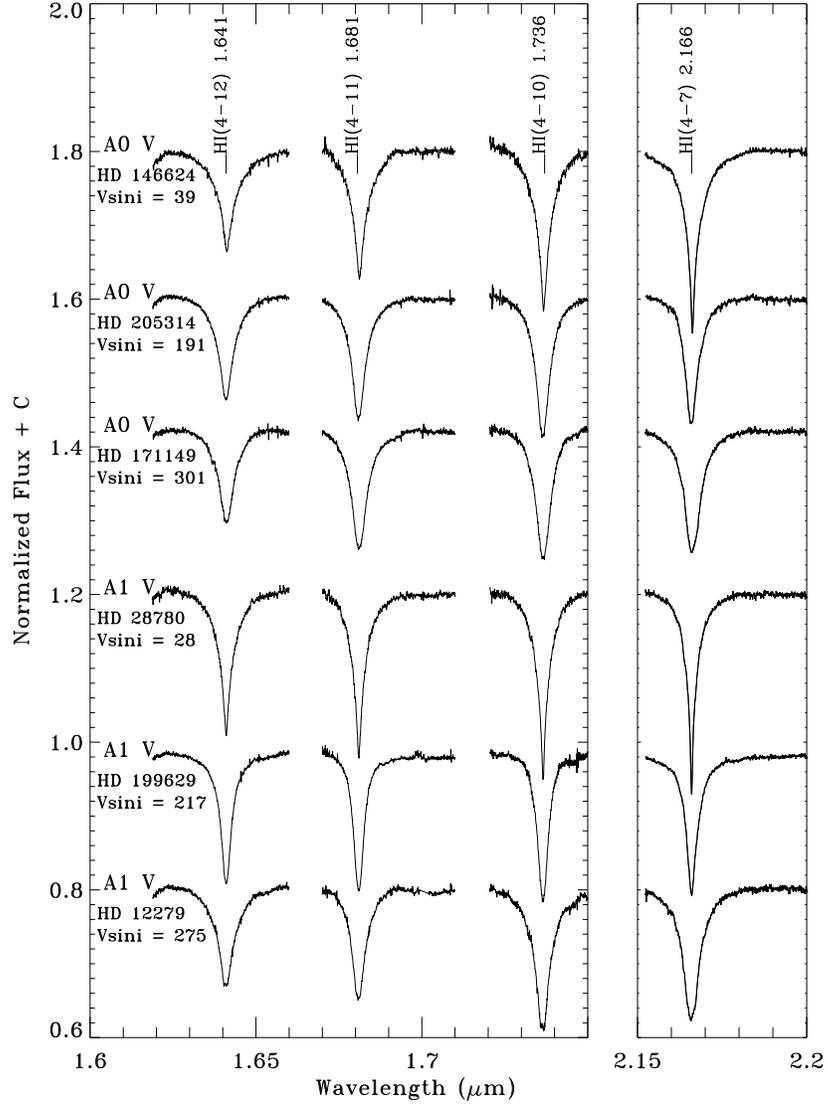}
\caption{{\bf A Dwarf stars useful for telluric standards.}  Note the significant
variation in profile characteristics with rotational velocity (velocities
are expressed in km/s).  High-order noise spikes, resulting from the telluric
corrections, have been artificially removed. \label{fig 14}}
\end{figure}

\end{document}